# Multi-trap optical tweezers based on Kinoform Silver Mean lenses

Francisco M. Muñoz-Pérez[1,2,*], Adrián Garmendía-Martínez[1], Vicente Ferrando[1], Juan C. Castro-Palacio[1], Walter D. Furlan[3] and Juan A. Monsoriu[1]

**Abstract:** In this paper, we present the design and implementation of multi-trap optical tweezers based on new quadrifocal kinoform lenses. The phase distribution of these diffractive lenses is characterized by the Silver Mean sequence. The focusing properties of the resulting aperiodic DOEs coined Kinoform Silver Mean Lenses (KSMLs) are numerically examined. It is shown that, under monochromatic illumination, a KSML drives most of the incoming light into four single foci whose focal lengths are related to the Silver ratio. In this way, a KSML improves the diffraction efficiency of binary Fresnel Silver Mean Zone Plates. Through experimental results, the simultaneous trapping of particles in the four focal planes and their three-dimensional manipulation is demonstrated.

## 1 Introduction

Laser manipulation has emerged as a pivotal tool in various interdisciplinary scientific fields, thanks to advancements in nanoscience. These applications span diverse domains like molecular biophysics, condensed matter physics, and quantum technologies [1–4]. Researchers have been actively developing trapping designs tailored for precise control of micro and nanostructures. These designs enable a wide range of applications, including atom cooling, particle confinement, thermodynamic studies of molecular motors in non-equilibrium states, and mechano-chemical analyses of nucleic acids, proteins, and viruses [5–10]. Optical tweezers are versatile instruments that play a crucial role in confining structures. Over the past decade, the use of structured beams has expanded the capabilities of optical tweezers. Techniques like elliptically polarized beams and vortex generation have enabled particle rotation, while interferometric and holographic methods allow simultaneous control of multiple particles [11–13]. The use of DOEs has also significantly enhanced the flexibility of beam-structure designs for optical tweezers [14]. Concurrently, there has been a resurgence of interest in DOEs because of their potential to revolutionize optical devices [15–18].

Whit in this context, diffractive lenses based on fractal sequences have attracted attention for their ability to generate multiple foci with extended depth of field [19–21]. It is also possible to design aperiodic diffractive lenses with interesting focusing and imaging properties by using Fibonacci [22], m-Bonacci [23], Thue-Morse [24], Precious Mean [25], and Walsh function [26], among others. Following this trend, our research group has recently proposed the Silver Mean Zone Plates (SMZPs) [27], which are binary-amplitude diffractive lenses constructed using the aperiodic Silver Mean sequence [28]. A SMZP is intrinsically quadrifocal producing four foci which focal lengths are related to the Pell numbers. This focal distribution is replicated along the optical axis at added fractions of the focal length of the equivalent Fresnel Zone Plate with the same number of zones. The use of the Silver Mean sequence in the design of diffractive lenses opens new possibilities for multi-trapping and particle manipulation.

[1]Centro de Tecnologías Físicas, Universitat Politècnica de València, 46022 València, Spain. [2]Laboratorio de Fibra Óptica, Universidad Politécnica de Tulancingo, División de Posgrado, Hidalgo, C.P. 43629, Mexico. [3]Departamento de Óptica y Optometría y Ciencias de la Visión, Universitat de València, Burjassot, E-46100, Spain.
*Email: fmmuope1@upvnet.upv.es

The combination of DOEs and optical tweezers holds great promise for advancing the field of particle manipulation and opening new avenues for scientific research [29], although the diffraction efficiency of these DOEs is crucial. Consequently, in this work, we introduce the concept of Kinoform Silver Mean Lenses (KMSLs), i.e., blazed zone plates with a sawtooth phase profile characterized by the silver mean sequence distributed on the square radial coordinate. As blazed DOEs, KSMLs drive most of the incoming light into the four main foci improving in this way the diffraction efficiency of binary SMZPs. That is why the implementation of KSMLs in the development of new multi-trap optical tweezers is a highly efficient option. We demonstrate experimentally that the focusing properties of KSMLs allow multiple trapping and three-dimensional particle manipulation at the four focal planes of the aperiodic diffractive lens.

## 2 Kinoform Silver Mean lenses design

The KSMLs considered in this work are blazed DOEs with a phase distribution based on the Silver mean sequence, which can be correlated to the Pell numbers. This sequence of numbers can be obtained from the recurrence relation $P_n = 2P_{n-1} + P_{n-2}$, for $n \geq 2$, being $P_0 = 0$ and $P_1 = 1$, so $P_n = \{0, 1, 2, 5, 12, 29, 70, ...\}$[30, 31]. The silver ratio is defined as the limit of the ratio between two consecutive Pell numbers:

$$\phi = \lim_{n \to \infty} \frac{P_n}{P_{n-1}} = 1 + \sqrt{2}. \tag{1}$$

Following a similar procedure based on Pell numbers, it is possible to generate the Silver Mean sequence starting from the seed elements $S_0 = \{B\}$ y $S_1 = \{A\}$. Then, the sequence of order $n \geq 2$ is constructed by applying the concatenation rule $S_n = \{S_{n-1} S_{n-1} S_{n-2}\}$. This aperiodic sequence can also be generated iteratively applying the substitution rules: $g(A) = \{AAB\}$ and $g(B) = \{A\}$ [28]. In this way, $S_2 = \{AAB\}$, $S_3 = \{AABAABA\}$, $S_4 = \{AABAABAAABAABAAAB\}$, etc. It is possible to observe that the total number of elements of a sequence of order $n$ is $P_n + P_{n-1}$, so it results from the sum of $P_n$ elements $A$ plus $P_{n-1}$ elements $B$.

In order to design a KSML of order $n$, we used a silver mean sequence, $S_n$, to define the phase distribution, $\Phi(\zeta)$, in the normalized square radial coordinate $\zeta = (r/a)^2$, where $r$ is the radial coordinate and $a$ is the lens radius. First we divide the interval $[0, 1]$ in $P_n + P_{n-1}$ sub-intervals of the same size. At each pair of sub-intervals $\{AB\}$, $\Phi(\zeta)$ is defined with a linear variation between $\Phi = 0$ rad and $\Phi = 2\pi$ rad, and $\Phi(\zeta) = 0$ rad otherwise (see Fig. 1(a)). In mathematical terms, the generating function of the radial phase for the n-$th$ order KSML can be written as,

$$\Phi_n(\zeta) = -\frac{2\pi}{2d_n} \sum_{j=1}^{P_{n-1}} rect\left[\frac{\zeta - \zeta_{n,j}}{2d_n}\right](d_m + \zeta_{n,j} - \zeta), \tag{2}$$

where $d_n = 1/(P_n + P_n - 1)$ and $\zeta_{n,j}$ is the position for the j-$th$ element B of the Pell sequence of order $n$. The phase pupil function of the KSML generated by Eq. 2 is represented in Fig. 1(b).

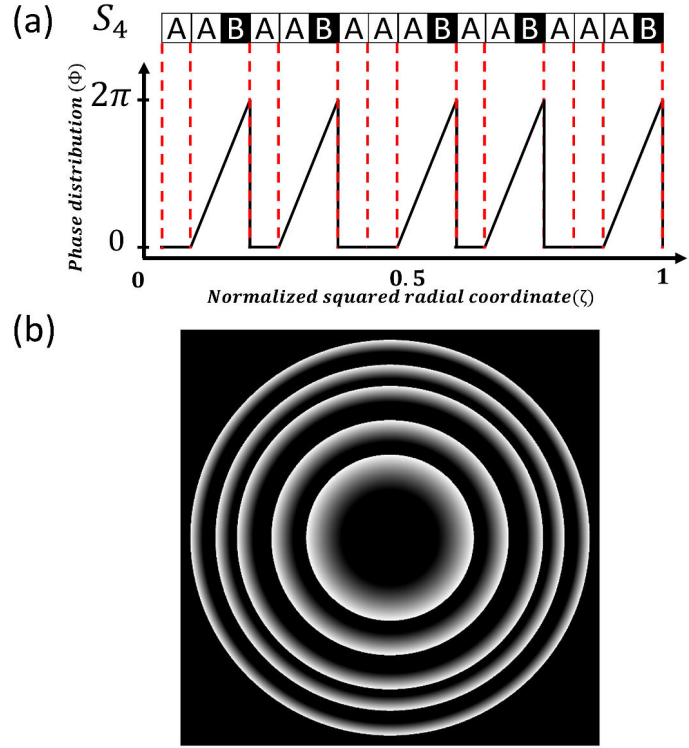

Figure 1: (a) Phase profile of a KSML of order $n = 4$ represented in the normalized square radial coordinate and (b) the corresponding phase pupil function in the radial coordinate after applying symmetry of revolution.

## 3 Focusing properties

To assess the focusing characteristics of KSMLs, we have calculated the axial irradiance under to monochromatic plane wave illumination, using the Fresnel approximation [32]

$$I(u) = 4\pi u^2 \left| \int_0^1 t(\zeta) \exp(-2\pi i u \zeta) d\zeta \right|^2, \tag{3}$$

where $u = a^2/2\lambda z$ is the reduced axial coordinate, $\lambda$ is the wavelength of the incident light, $z$ is the axial distance, and $t(\zeta) = \exp[-i\Phi(\zeta)]$ is the transmittance function, being $\Phi(\zeta)$ the phase function of the lens given by Eq. 2. Figure 2 shows the axial irradiance distribution computed for KSMLs of orders $n = 4, 5,$ and $6$. The KSMLs provide four single foci distributed axially, which focal lengths can be correlated with the Silver ratio. As can be seen, the positions of the focal points change as the order $n$ increases, causing the distance between the resulting focal planes also to increase. For the three lenses herein studied ($a = 4.2$ mm, $\lambda = 1064$ nm), the focal planes are located at $F_a = 3.310$ D, $F_b = 4.676$ D, $F_c = 6.611$ D and $F_d = 7.974$ D for $S = 6$, $F_a = 1.376$ D, $F_b = 1.936$ D, $F_c = 2.732$ D and $F_d = 3.309$ D for $S = 5$, and $F_a = 0.584$ D, $F_b = 0.808$ D, $F_c = 1.124$ D and $F_d = 1.380$ D for $S = 4$. Like in binary SMZPs, it is possible to derive the ratio between the focal distances as $\frac{F_d}{F_a} \approx 1 + \sqrt{2}$, $\frac{F_d}{F_b} = 2$, and $\frac{F_c}{F_b} \approx 1 + \frac{1}{\phi} = \sqrt{2}$. If we compute these ratios with the focal lengths obtained numerically for the KSMLs, they are considerably closer to the expected ones.

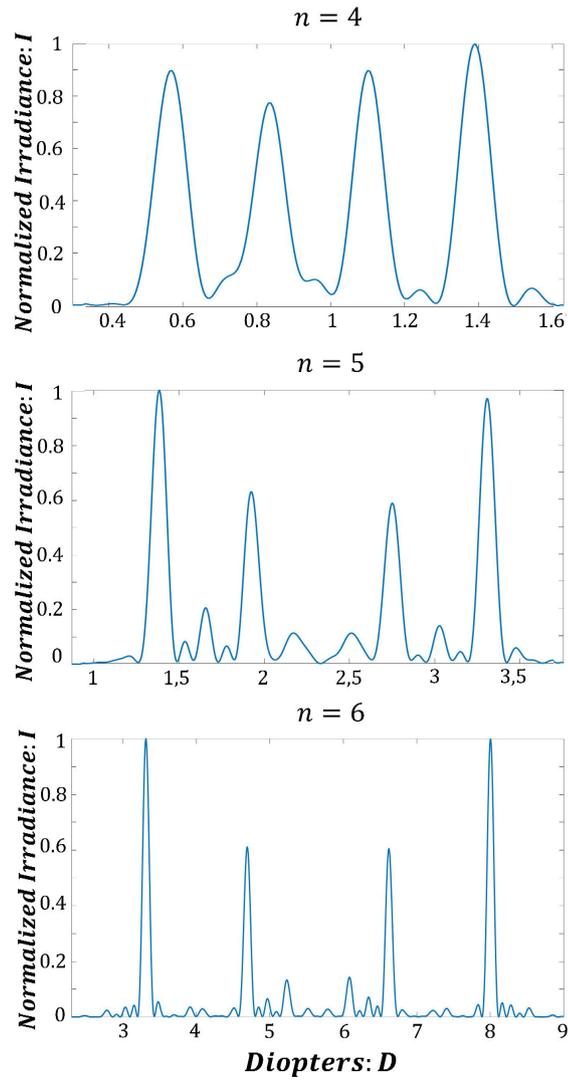

Figure 2: Normalized axial irradiance generated by KSMLs of three different orders $n = 4$, $n = 5$, and $n = 6$.

## 4 Experimental results

We implemented an optical tweezers setup to show the multi-trapping capabilities of KSMLs, with a diagram depicted in Fig. 3. A beam is emitted from a CW laser ($\lambda = 1064$ nm, Laser Quantum, Mod. Opus 1064) with a maximum power of 3 W. A half-wave plate ($\lambda/2$) is placed at the laser output followed by a linear polarizer ($P$), which changes the direction of the beam linear polarization. The laser beam is then redirected by mirrors (M1 and M2) and expanded through a system of magnification 3, formed by lenses L1 and L2 (focal length $f_1 = 50$ mm and $f_2 = 150$ mm). The KSML was projected on a spatial light modulator (SLM) (Holoeye PLUTO-2.1-NIR-149, phase-type, pixel size 8 $\mu m$ and resolution 1920 x 1080 pixels) screen. The SLM is configured for a $2.1\pi$ phase at a wavelength $\lambda = 1064$ nm. The resulting KSML beam, as modulated at the SLM, is directed by a $4f$ system formed by L3 ($f_3 = 150$ mm) and L4 (focal length $f_4 = 150$ mm). A 1D blazed gratings are incorporated into each KSML for the purpose of serving as linear phase carriers. Their role is to direct diffracted light towards the first order of diffraction, effectively eliminating noise caused by specular reflection from higher diffraction orders.

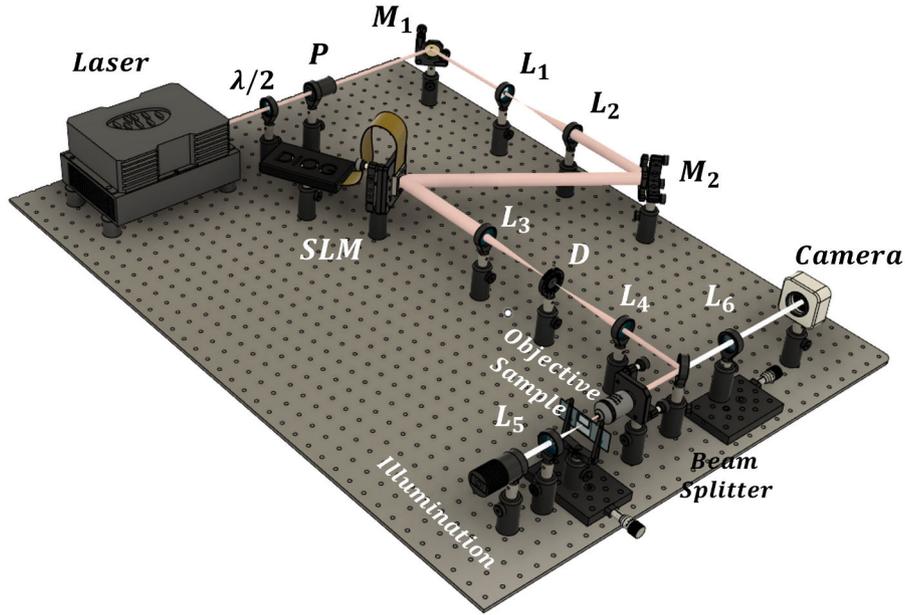

Figure 3: Experimental setup based on KSMLs for multi-trapping and manipulation of particles.

To achieve this, the beam undergoes spatial filtering through a diaphragm (denoted as D) positioned at the focal point of lens L3, allowing only the first order of diffraction to pass through. To ensure alignment, the added linear phase carrier is slightly tilted by adjusting the Spatial Light Modulator (SLM) so that the first order of diffraction aligns with the optical axis of the diaphragm. Subsequently, the KSML image passes through a high-numerical aperture oil-immersion objective (Olympus UPLFLN 100X, NA= 1.3) located at the focal plane of L4. An LED light source (Thorlabs, Mounted High-Power, 1300 mA, Mod. MCWHL7) is employed to illuminate the sample. The light emitted from the LED is collimated and then focused onto the sample using lens L5 (with a focal length of 30 mm). With the purpose of capturing images, a Beam Splitter (BS) is utilized to transmit visible light from the sample while preventing reflections of infrared light from reaching the imaging system. The image is then focused using lens L6 (with a focal length of 50 mm). Finally, a CMOS camera sensor (Edmund Optics, Mod. EO-10012C) is employed to capture the images generated in this optical configuration.

By using Eq. 3 we have computed the axial irradiance produced by a KSML of order $n = 4$ and, for comparison purposes, those corresponding to binary-amplitude and binary-phase SMZPs of the some order. The results are shown in Fig. 4 (left panel). It is important to mention that the high numerical aperture lens performs a rescaling of the focal points, that is, the power of the objective ($P_O = 555$ D or $\sim 150$ D in distilled water) modifies the axial distances of the foci. As can be seen, the KSML drives most of the incoming light into four single foci corresponding to the first diffraction order of the lens. On the other hand, SMZPs provide multiple diffraction orders due to the binary nature of the structure. Note that the higher orders also present four diffraction peaks. The first order foci of the KSML coincide with the foci of the binary-phase and binary-amplitude SMZPs, but their relative intensity is approximately 60% and 90% lower, respectively. Due to the low efficiency of binary phase and amplitude plates, the generation of stable optical multi-traps from these lenses is difficult. That is the reason why KSMLs are a more efficient option in the development of multi-trap optical tweezers.

We next analyze the experimental results that demonstrate the multi-trapping of microparticles in the focal planes of KSMLs. Figure 4 shows the comparison between the experimental trapping position of each microparticle and the numerically expected positions. It is possible to observe (right panel) the stable trapping of four polystyrene microparticles (diameter $\sim 2\mu m$) using a KSML. The small angle at which the beam is incident causes a relocation of the axial positions in the transverse plane, which simplifies the simultaneous observation of particles located in separate planes. Each particle is trapped at each focal point, so the multi-focusing capability of a KSML allows multiple trapping of microparticles. As is well known, the trapping dynamics cause the particles to undergo a slight displacement of the optical trap, which causes a small variation in the expected position of the focal points [33]. Nevertheless, as can be seen in Fig.4, the relative positions of the trapped microparticles are similar to the positions of the focal points obtained numerically.

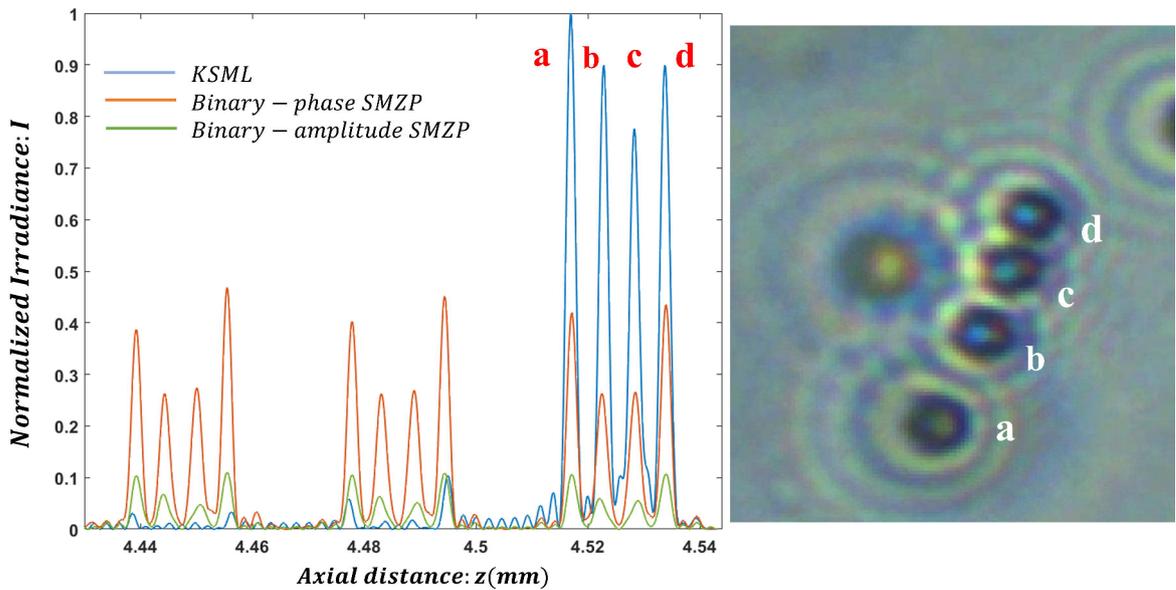

Figure 4: Particle multi-trapping with optical tweezers based on a KSML of order $n = 4$. Comparison of the axial position of the focal planes obtained numerically (left panel) and the experimental relative position of the trapped particles (right panel)

# 5 Conclusions

A new multifocal kinoform lens based on the Silver Mean sequence and a multi-trap optical tweezers for multiple axial captures of microparticles are presented. The KSML has the characteristic of forming four single focal planes, improving the diffraction efficiency of binary SMZPs. It is also shown that the ratio between the focal lengths are related to the Silver ratio, obtaining a very good agreement. The multifocal feature of KSMLs allows its application in multiple axial capture, creating an alternative way to three-dimensional manipulation. Unlike the binary SMPZ, the high efficiency of the KSML lens makes it an option in the generation of multiple optical traps. The experimental position of particles trapped in each focal plane is compared with the numerically obtained focal plane positions. Positioning particles along a line at controlled distances would allow the interactions between them to be explored under laser irradiation. Coupled with its application in a system of optical tweezers, we consider that this type of aperiodic lens will allow the generation of multiple applications in various fields, such as ophthalmology, microscopy, or quantum computing.

# Funding

This work was supported by the Spanish Ministerio de Ciencia e Innovación (grant PID2022-142407NB-I00) and by Generalitat Valenciana (grant CIPROM/2022/30), Spain. F.M.M.P. and A.G.M. acknowledge the financial support from the Universitat Politècnica de València (PAID-01-20-25) and from the Generalitat Valenciana (GRISOLIAP/2021/121), respectively.

# Disclosures

The authors declare no conflicts of interest.